\documentclass{aa}

\usepackage{graphicx}
\begin{document}

   \title{An infrared study of local galaxy mergers}

   \author{A. Carpineti\inst{1}, S. Kaviraj\inst{2,1,3}, A. K. Hyde\inst{1}, D. L. Clements\inst{1}, K. Schawinski\inst{5}, D. Darg\inst{3}, C. J. Lintott\inst{3,4}
          }

   \institute{Blackett Laboratory, Imperial College London\\
              \email{alfredo.carpineti07@imperial.ac.uk}
              \and
              University of Hertfordshire, Hatfield, Hertfordshire, AL10 9AB UK
         \and
             Department of Physics, University of Oxford, Keble Road, Oxford OX1 3RH UK
\and
Adler Planetarium, 1300 S. Lakeshore Drive, Chicago, IL 60605.
\and
Institute for Astronomy, Department of Physics, ETH Zurich, Wolfgang-Pauli-Strasse 16, CH-8093 Zurich, Switzerland
             }

\date{}


  \abstract {We combine a large, homogeneous sample of $\sim$3000 local mergers with the Imperial IRAS Faint Source Redshift Catalogue (IIFSCz), to perform a blind far-infrared (FIR) study of the local merger population. The IRAS-detected mergers are mostly ($98\%$) spiral-spiral systems, residing in low density environments, a median FIR luminosity of $10^{11} L_\odot$ (which translates to a median star formation rate of around 15$M_\odot yr^{-1}$). The FIR luminosity -- and therefore the star formation rate -- shows little correlation with group richness and scales with the total stellar mass of the system, with little or no dependence on the merger mass ratio. In particular, minor mergers (mass ratios  $<1:3$) are capable of driving strong star formation (between 10 and $173 M_\odot yr^{-1}$) and producing systems that are classified as LIRGs, luminous infrared galaxies ($65\%$ of our LIRGs are minor mergers), with some minor-merging systems being close to the ultra luminous infrared galaxy (ULIRG) limit. Optical emission line ratios indicate that the AGN fraction increases with increasing FIR luminosity, with all ULIRG mergers having some form of AGN activity. Finally, we estimate that the LIRG-to-ULIRG transition along a merger sequence typically takes place over a relatively short timescale of $\sim$160 Myr.}


\authorrunning{Carpineti et al.}
\maketitle


\section{Introduction}
{Mergers are a fundamental feature of our current understanding of
the Universe: the standard $\Lambda$CDM cosmology \citep{Blu84,
Fre01,Efs02, Pry02, Spe07} is based on a hierarchical structure
formation paradigm, in which smaller systems merge to form
progressively larger ones \citep{Whi78,Sea78}. Galaxy merging is
expected to drive strong star formation episodes \citep{Bar91,
Mih96}, contribute to and regulate the growth of black holes
\citep{Kau03, Car05, Spr05, Sch09}, and produce morphological
transformations \citep{Too77, Bar96}. Given the essential role of
mergers in the evolution of the visible Universe, it is important
to extend our knowledge of their characteristics beyond the
information that we can gather by studying mergers in UV/optical
wavelengths alone.

Ultraviolet/optical astronomy, while the cornerstone of our astronomical
knowledge, does not provide us with a complete picture of star
formation. Short-wavelength photons are affected by absorption and
scattering by cold interstellar dust, causing a dimming in the
UV/optical light and an apparent reddening of the source
\citep{Dra03, Zub04, Dra09}. Indeed, around $50\%$ of all the
energy produced by star formation and {active galactic nuclei}
(AGNs) over cosmic time has been absorbed by molecular clouds and
re-emitted in the {far-infrared} wavebands \citep[$3.5 - 1000
\mu$m, e.g.][]{Pug96,Fix98,Pei99, Hau01, Dol06}. Obscuration could
be more severe in mergers because of their dustier cores, making it
challenging to measure the total star formation activity via the
UV/optical wavelengths \citep{Spi78, Ken98, Pei99}. However, while
UV/optical signatures are likely to be affected, the star
formation activity can be better studied using far-infrared (FIR)
wavelengths, as the peak of emission from cold dust in the star-forming regions lies in the FIR \citep[e.g.][]{Shu87, Pei99,
Cha01}.

In recent decades, several studies have looked at the connection
between mergers and FIR-bright galaxies. Since the discovery of
luminous and ultra-luminous infrared galaxies (LIRGs and ULIRGs),
which exhibit FIR luminosities exceeding $10^{11}\text{L}_{\odot}$
and $10^{12}\text{L}_{\odot}$, respectively \citep[e.g.][]{Soi84,
San96}, studies have routinely found evidence for morphological
disturbances or ongoing mergers in FIR-bright galaxies
\citep{Kle87,San87,Hut87,Vad87,San88,San88b, San96, Cle96, Hop06,
You09, Hwa10}. While most of these studies have started with a
sample of FIR-bright galaxies and found evidence for merger
activity within some of them, we approach the infrared properties
of mergers from the opposite angle.

\begin{figure*}
\centering
\begin{tabular}{c c}
\includegraphics[width=90mm]{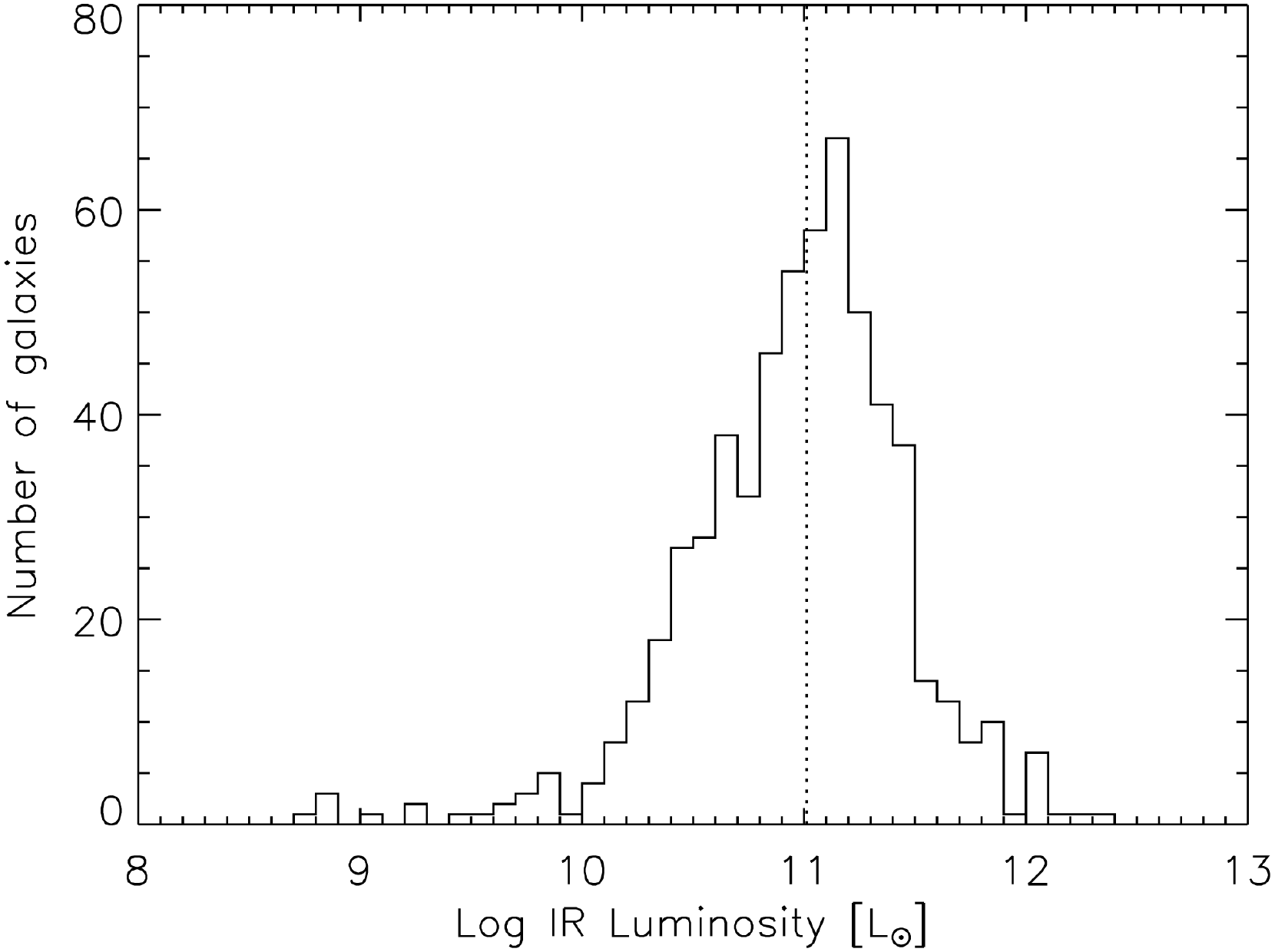} & \includegraphics[width=90mm]{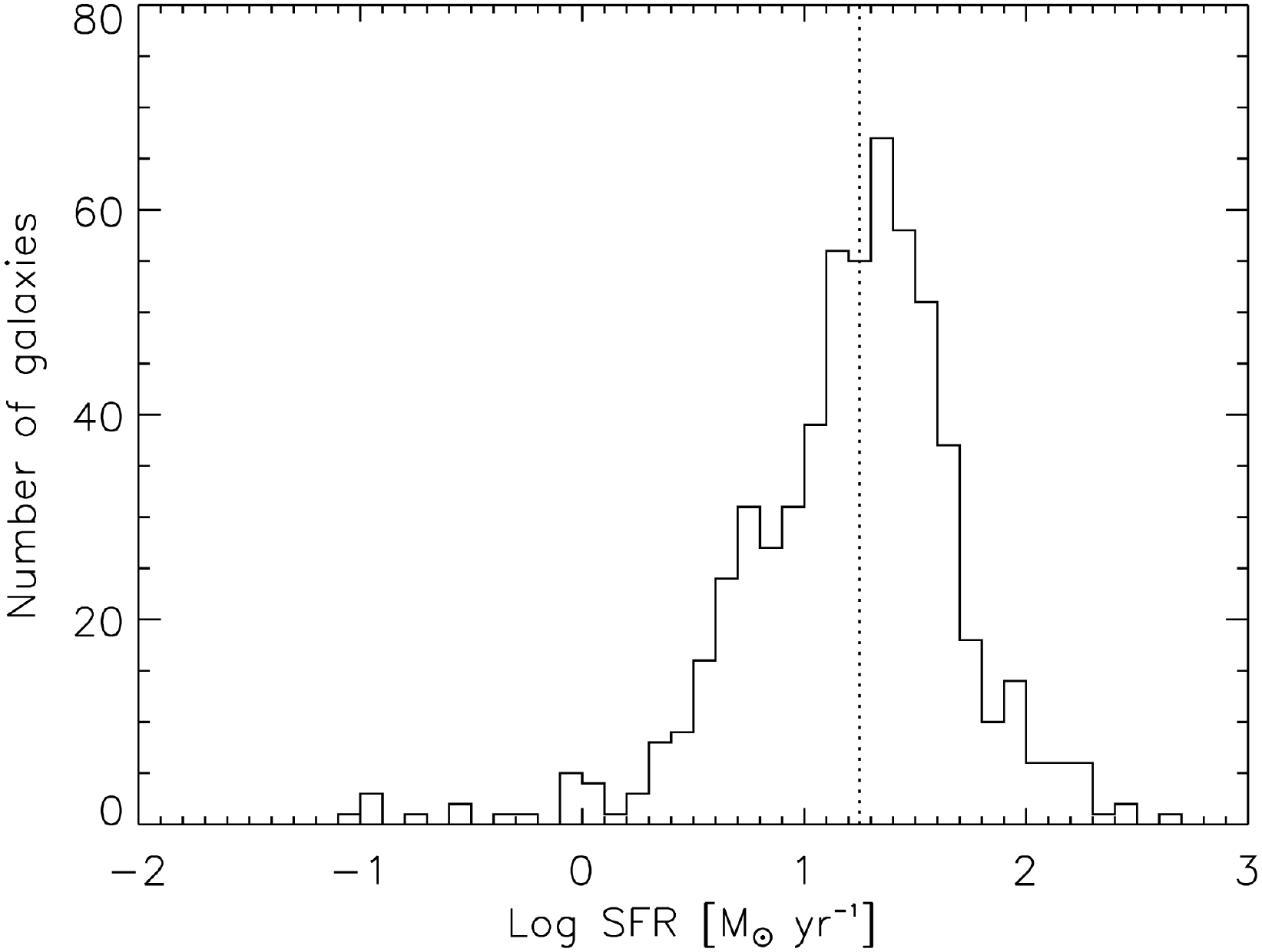}\\
\end{tabular}
\caption{\small Distribution of the FIR (8-1000 $\mu$m)
luminosities of the IRAS-detected galaxy mergers (left) and the
corresponding star formation rates (right), following Eqn 4 in
Kennicutt (1998).}\label{fig:LogIR3}
\end{figure*}

\begin{figure}
\centering
\includegraphics[width=90mm]{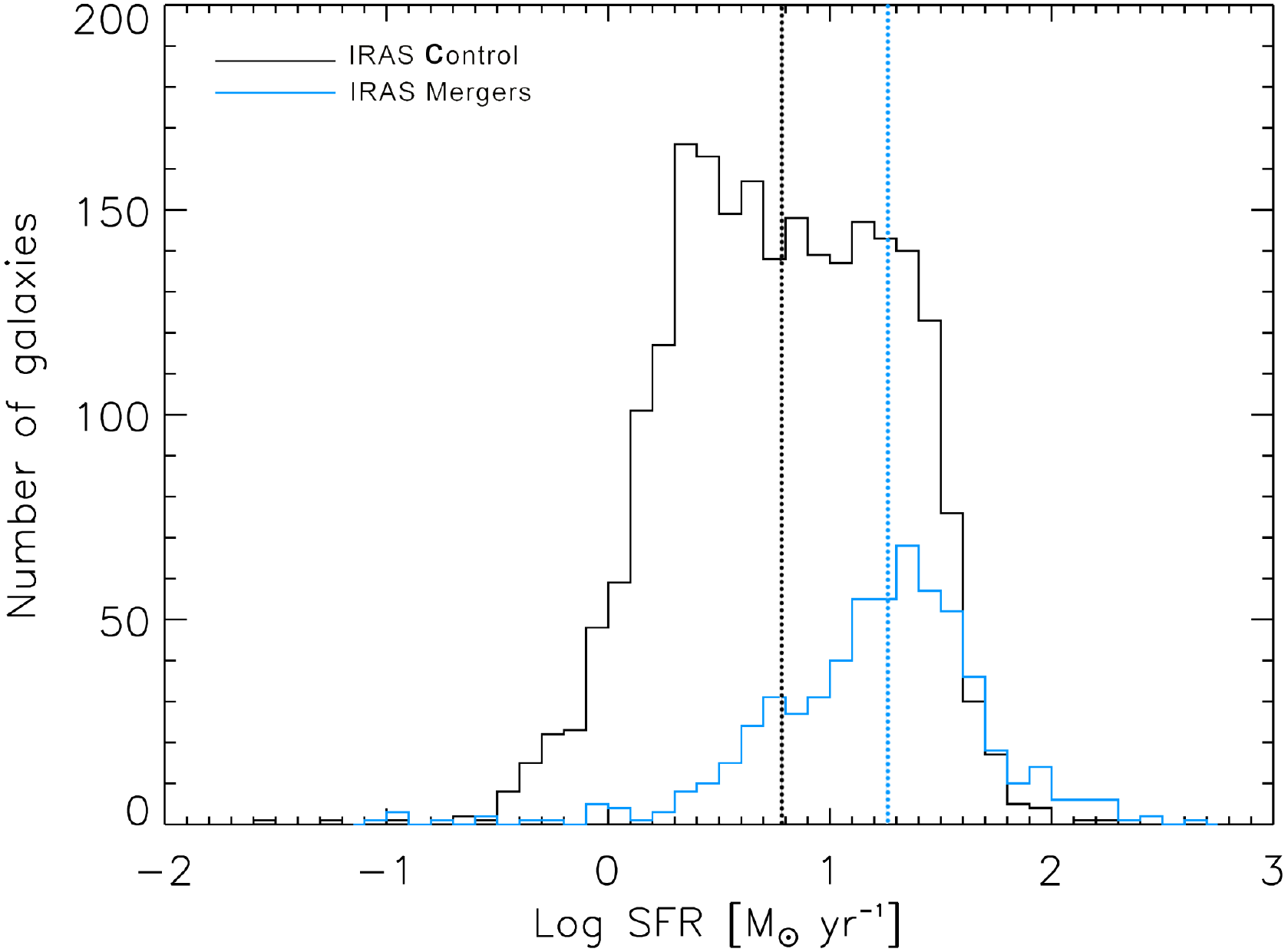}
\caption{\small Distribution of star formation rates for the
IRAS-detected mergers and for a control sample of non-merging
galaxies detected by IRAS. Median values are indicated using the
dotted vertical lines.}\label{fig:LogSFR3}
\end{figure}

We start with a large, visually selected sample of local galaxy
mergers $(z < 0.1)$ and then explore their infrared properties via
their IRAS photometry, thus performing a large \emph{blind}
statistical study of nearby mergers in the FIR. Such a study has
been challenging in the past because large unbiased samples of
true mergers in the local Universe have been lacking
\citep{Dar09a, Dar09b}}. Notwithstanding IRAS' low sensitivity
compared to newer instruments (e.g. Herschel), an IRAS-based
analysis is useful here because the large IRAS beam size of
1.5'$\times$4.7' (at 60$\mu$m, \citealp{IRAS, Neu84}) allows us to
study the entire merging system as a single source. The SDSS
mergers are isolated systems, so contamination by nearby sources
within the IRAS beam is negligible. \\

The plan for this paper is as follows. In Section 2, we discuss
the general properties of the merger sample that underpins this
study. In Section 3, we study how the star formation in mergers
depends on the total mass and mass ratio of the merging systems,
while in Section 4 we study the impact of local environment on
star formation in our mergers. In Section 5, we explore the
emission line activity in our sample and in Section 6, we explore
the timescale for LIRGs to transform into ULIRGs along the merger
sequence. We summarize our results in Section 7. Throughout this
paper we adopt the standard cosmological parameters from
\cite{Spe07} ($H_0=70kms^{-1}Mpc^{-1}$, $\Omega_\Lambda=0.73$,
$\Omega_M=0.27$).


\section{The merger sample and basic properties}
Our sample of visually classified mergers was produced via the
Galaxy Zoo (GZ) project \citep{Lin08}. GZ is uniquely powerful in
detecting rare classes of objects like mergers, which can only be
reliably selected via direct visual inspection of galaxy images.
GZ had enlisted over 500,000 volunteers from the general public to
morphologically classify (as spiral, elliptical and mergers), through visual inspection, the entire
SDSS spectroscopic sample {\citep{Yor00, Ade08}}. {This preliminary classification was used by Darg et al. (2010b) to build a merger catalogue of over three thousand objects. The raw parameter $f_m$, called the weighted-merger-vote fraction, is used to select merging galaxies from the whole sample.  $f_m$ simply quantifies the probability that a certain image was in fact the image of a merger based on all the classifications of that particular image. The parameter $f_m$ ranges from 0 to 1 so that an object with $f_m = 0 $ should look nothing like a merger and $f_m = 1$ should look unmistakably so. To find mergers within the GZ catalogue, a cut of $0.4 <f_m <1.0 $ was applied. The reason behind the cut on the weighted-merger-vote fraction was the high occurrence of false positives (which are virtually non-existent in the interval $f_m>0.6 $). A second layer of visual inspection, performed by the team, was used to remove any non-merging systems, visually select an appropriate SDSS object to represent the merging partner, and assign morphologies to the galaxies in each merging system. 
}

\begin{figure*}
\centering
\begin{tabular}{c c}
 \includegraphics[width=85mm]{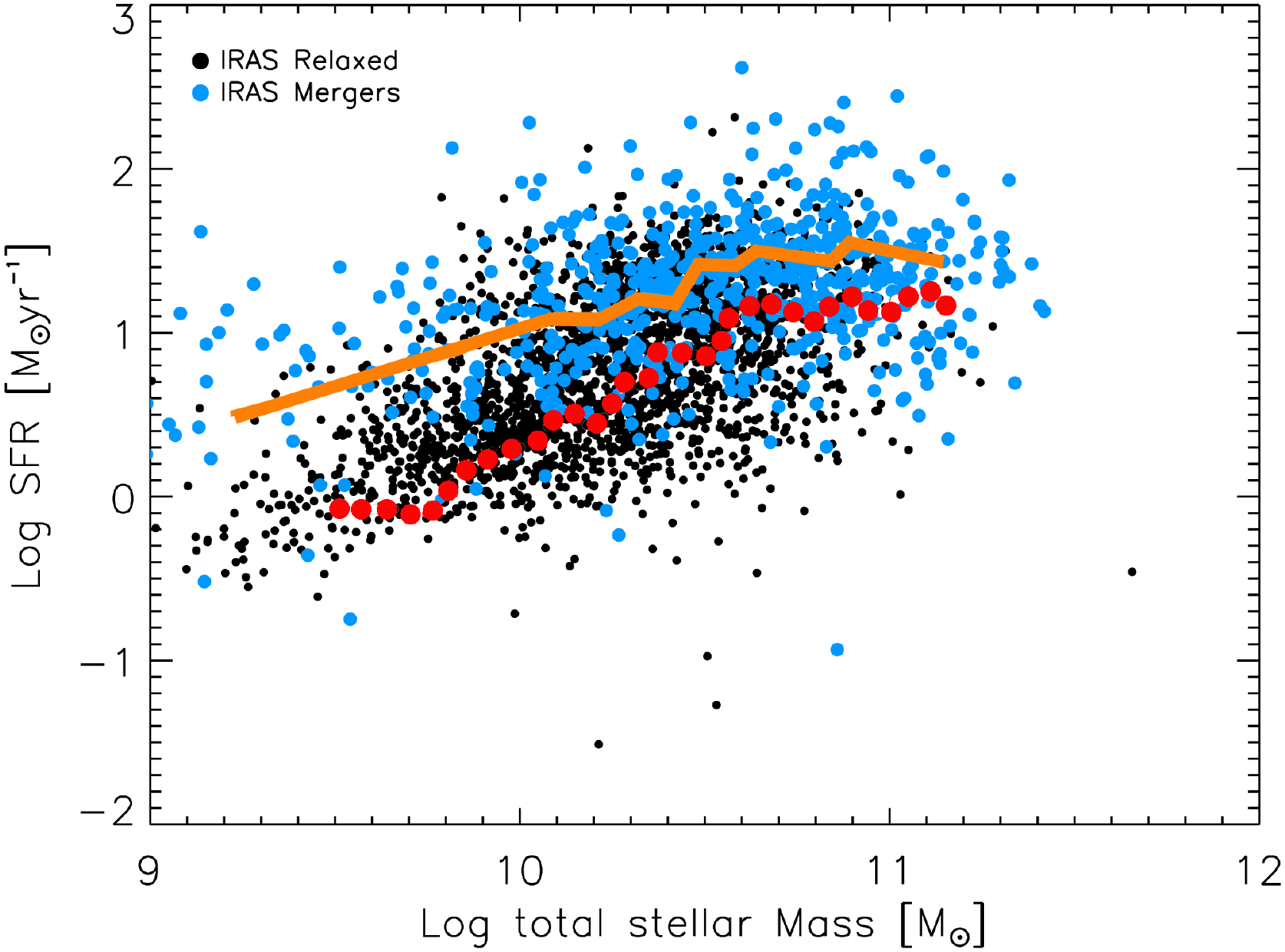} &
 \includegraphics[width=95mm]{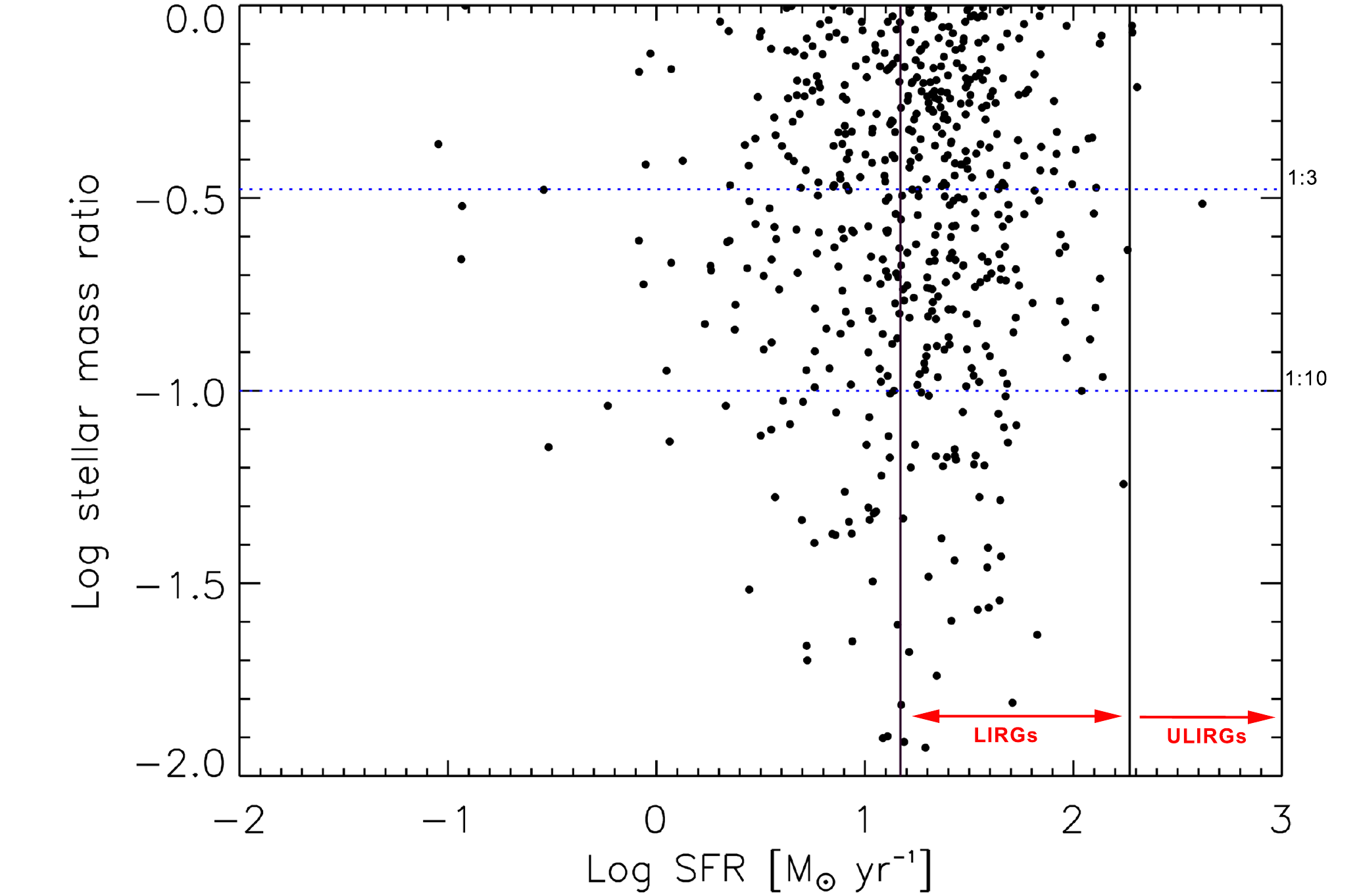}
\end{tabular}
\caption{\small LEFT: The star formation rate versus the total stellar
mass for mergers and the control sample {of relaxed non-merger galaxies.}. RIGHT: The star formation
rate in mergers plotted against the merger mass ratio. Mass ratios
of 1:3 and 1:10 are indicated using the dotted horizontal lines.
From left to right the solid vertical lines represent the star
formation rates for LIRGs and ULIRGs. There is no significant
trend between SFR and mass ratio. It is also important to note
that minor mergers play a significant role in the formation of
LIRGs. The left plot also shows a progressive one-sigma fit
to the mergers (solid orange) and control sample (dashed
red).}\label{fig:ratio3}
\end{figure*}

 The
final sample of GZ merger pairs (on which this study is based) contains
3373 objects, unbiased in morphology and local environment and
with mass ratios typically between 1:1 and 1:10 at $z<0.07$. {The separation range for these galaxies goes up to 100 kpc. Since the merger classification is based on morphological disturbances, the separation peaks at 10 kpc and the number of objects with a separation greater than 20 kpc quickly dwindles. Thus a significant proportion of very early stage mergers are excluded from the catalogue.} We refer readers to \citet{Dar09a} for a complete description of the merger sample.

The GZ merger sample is cross-matched with the Imperial IRAS-FSC
Redshift Catalogue (IIFSCz; \citealp{Wan09}), a sample of
$\sim$60,000 galaxies selected at 60 $\mu$m from the IRAS Faint
Source Catalogue (FSC; \citealp{Mos92}). {The IIFSCz
catalogue provides FIR fluxes based on the best-fitting infrared
templates of \citet{RR08}. For the flux limit at $60\mu$m
($f(60)\sim$ 0.36 Jy), 90 per cent of sources have spectroscopic
redshifts from the SDSS.}

Cross-matching the GZ mergers with the IIFSCz yields 606 systems.
Our analysis therefore applies to the brightest 18\% of the
FIR-detected GZ mergers. In this merger sample, 594 mergers are
spiral-spiral merging pairs, with only 12 having an elliptical
progenitor. There are 274 LIRGs (45\%) and 10 ULIRGs (2\%) in our
sample, with the ULIRGs typically being nearly-coalesced, i.e. in
the very final stages of the merger. we note that the Darg et al.
merger catalogue is biased \emph{against} ULIRGs, since
they typically do not show two merging nuclei with tidal bridges
between them, which are the main criteria for a galaxy pair to be
defined as a merger in GZ (e.g. the ULIRG fraction from the GOALS survey is $10\%$, \citealp{Arm09}).

The $L_{FIR}$ in our IRAS-detected mergers ranges from around
$10^{9} \text{L}_\odot$ to over $10^{12.3} \text{L}_\odot$, with a
median of $\sim10^{11} \text{L}_\odot$, which corresponds to a
star formation rate (SFR) of $15 \text{M}_\odot\; \text{yr}^{-1}$
(see Figure \ref{fig:LogIR3}), calculated using \citet{Ken98}:

\begin{equation}
SFR_{FIR}(\text{M}_\odot
\text{yr}^{-1})=4.5\times10^{-44}\text{L}_{FIR}(\text{ergs
}\text{s}^{-1}). \label{SFR-FIR 3}
\end{equation}

To compare the properties of our IRAS-detected mergers to the
general galaxy population in our subsequent analysis, we construct
a control sample of 2300 relaxed (i.e. non-merging) galaxies from Galaxy Zoo and
the IIFSCz, selected to have the same redshift and $r$-band magnitude ranges as the mergers. {Most of the galaxies in our sample are spirals (2250) with a small fraction of ellipticals (50) to keep a similar proportion to the morphology in the IRAS mergers (49.5 spirals per elliptical).}


\section{Dependence of SFR on system mass and mass ratio}
We begin by studying how the SFRs of mergers are influenced by
their total stellar mass and the merger mass ratio. {The stellar mass was calculated by \cite{Dar09a} by fitting each galaxy's SDSS photometry  to a library of photometries produced by a variety of two-component star formation histories, where the older burst is taken as a simple stellar population while the more recent burst is modelled by an exponential to take into account the extended star formation history of mergers. The library of SEDs are generated using the \cite{Mara98,Mara05} stellar models. Both components have stellar populations with variable age with fixed solar metallicity and Salpeter IMF \citep{Sal55}. Dust is implemented using a \cite{Cal00} law  with an E(B-V) varying between 0 and 0.6. The uncertainty on the masses for the GZ merger sample is 0.2 dex \citep{Dar09b, Dar09a}}

In
Figure \ref{fig:ratio3} we study the dependence of the SFR on the
total (stellar) mass of the system (left panel) and on the merger
mass ratio (right panel).

{The SFR is positively correlated with the total stellar mass of
the system, with LIRGs and ULIRGs residing mostly at the higher
end of the mass spectrum. While the positive correlation exists
both for the mergers and the control sample, the mergers show a
largely mass-independent {enhancement in SFR of {0.5 dex}.} The enhancement at the low mass end of the distribution is within the one-sigma error in the fit. There is a wide spread in the SFR of the mergers at the low mass end and a proportionally smaller number density compared to the higher end of the distribution. The linear fit for the distributions are
\begin{equation}
\log(SFR)=0.5(\pm1.2)\log(Mass)-4.1 
\end{equation}
for the mergers, and
\begin{equation}
\log(SFR)=0.8(\pm1.3)log(Mass)-8.1
\end{equation}
for the control sample. The control sample fit is consistent with the SFR-Mass relation for SDSS galaxies found in \citet{Elb07} and in \citet{Will15}, which also reported a SFR enhancement in the local merger population.}

We find no apparent dependence between the mass ratio of the
merger and the SFR in the merging galaxies. 
As an ulterior check we compare merger systems within the same total stellar mass range: we plot the mass ratio versus the total stellar mass and the mass ratio versus the specific SFR. No correlation was found in those plots and they hence have not been included.
Minor mergers (those with mass ratios less than 1:3) seem capable of
producing IR-bright systems, including LIRGs and systems that are
close to the ULIRG limit ($L>10^{12}l_{\odot}$). Our analysis,
therefore, suggests that major mergers are not the only process
that can trigger strong star formation episodes (between 10 and $173 M_\odot yr^{-1}$), somewhat in
contradiction with the wider literature \citep{Car05,Spr05,Cox06}.
In particular, minor mergers appear to play an influential
role ($65\%$ of LIRGs are minor mergers) in triggering such strong star formation episodes, { in
agreement with the findings of recent observational work \citep{Scu12, Kav14b, Kav14}.}

Overall, we find that the star formation activity in mergers is
correlated more strongly with the total stellar mass of the
merging system, with the system mass ratio being of negligible
importance.



\begin{figure}
\centering
\begin{tabular}{c}
\includegraphics[width=90mm]{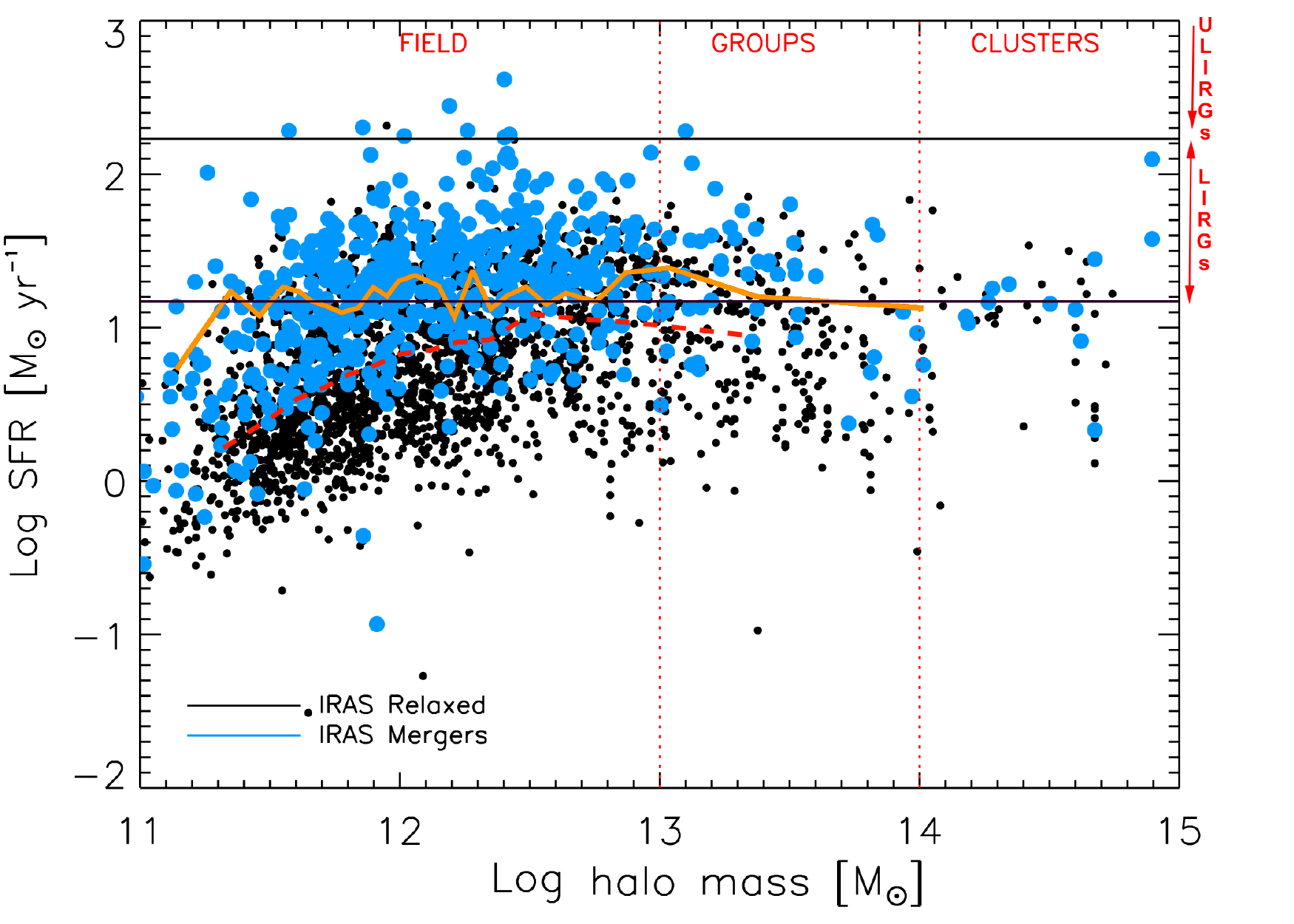}
\end{tabular}
\caption{\small The star formation rate (SFR) of the IRAS-detected
mergers (blue) and the LTG control sample (black) plotted against
local environment. The plot also shows a progressive one-sigma fit
to the mergers (solid orange) and control sample (dashed
red).}\label{fig:Rho3}
\end{figure}

\section{Dependence of SFR on local environment}
In Figure \ref{fig:Rho3} we explore the dependence of the merger
SFR on local environment. We use the Yang et al. environment
catalogue \citep{Yang07} to estimate the environment of our
merging systems. Yang et al. estimate the dark matter halo mass of
individual galaxies, which can be used as a proxy for the local
environment of the galaxy. Typically, haloes with masses lower than
10$^{13}$ M$_{\odot}$ are considered to be the field, haloes
with masses between 10$^{13}$ M$_{\odot}$ and 10$^{14}$
M$_{\odot}$ are considered to be in group-like environments,
while haloes with masses greater than 10$^{14}$ M$_{\odot}$ host
clusters \citep{Kav09}.

Across the range of environments sampled by our mergers, we do not
find a strong correlation between the SFR in merging galaxies and
their local environment. The relationship between SFR and
environment in the mergers is similar to that in the control
sample, except that the mergers show an enhancement in SFR of
around 0.5 dex, largely irrespective of environment. In agreement
with previous studies \citep[e.g.][]{Hwa10, Dar09a, Car12}, we
find that, like the general merger population, the IRAS-detected
mergers tend to favour lower-density environments. Most of the
IRAS-detected mergers are found in groups and the field (95\% of
the sample favours such low-density environments), with ULIRGs
exclusively inhabiting the field.


\section{Emission line activity}
{We use the nuclear activity classification from \cite{Dar09a} to investigate the ionization mechanisms in the
IRAS-detected mergers and probe the connection between AGN activity and FIR luminosity. The classification was done by performing a BPT analysis  \citep{BPT81} on the full GZ merger catalogue, using optical emission-line ratios from the SDSS. The line ratios used were $[O III]/H\beta$, $[N II]/H\alpha$, $[S II]/H\alpha$ and $[O I]/H\alpha$, to obtain three reddening insensitive diagnostic diagrams following  \cite{Vei87}. \cite{Dar09a} used the demarcation lines from \cite{Kau03,Kew06}) to
separate galaxies into ones that are star-forming, AGN,
LINERs, or SF/AGN (i.e. contain both star formation and
AGN), using a signal-to-noise (S/N) threshold of 3 in all lines.
Galaxies which do not have S/N $>$ 3 in $H\alpha, H\beta, NII, OIII$ lines are
classified as quiescent. The ratios were computed using the
GANDALF code \citep{Sar06}. We used the same approach to classify the ionization mechanisms for the control sample.}

\begin{figure*}
\centering
\includegraphics[width=180mm]{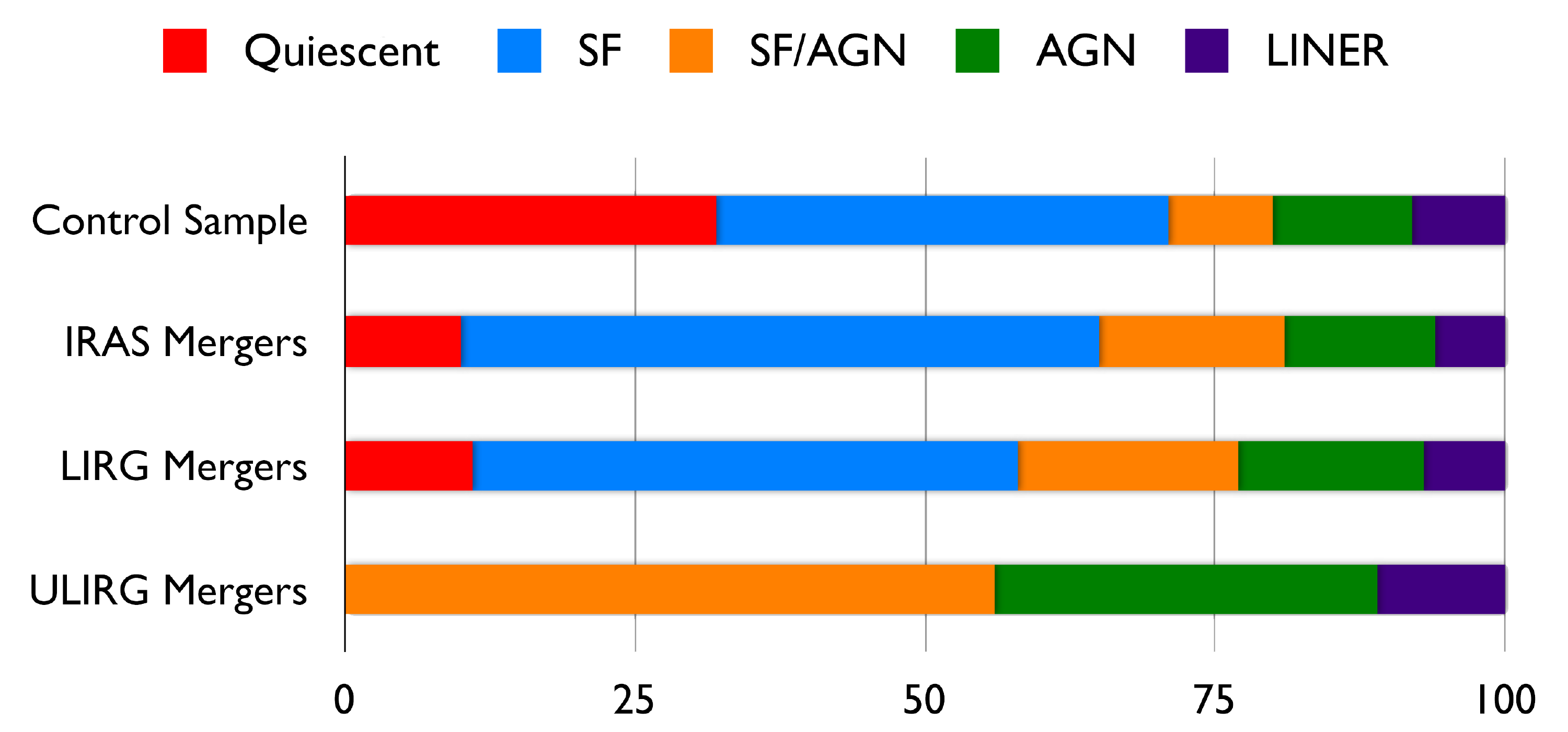}
\caption{\small Emission-line classes for the control sample,
IRAS-detected mergers and the subsets of the IRAS-detected mergers
that are classified as LIRGs and ULIRGs.}\label{fig:Comp3}
\end{figure*}

In Figure \ref{fig:Comp3}, we split the IRAS-detected mergers into
these emission-line categories and study how they change with FIR
luminosity. We also perform an equivalent analysis for our control
sample of galaxies from the IIFSCz. The {majority ($71\%$) of the
IRAS-detected mergers  are classified as star-forming, with a
significant minority ($35\%$)} hosting an AGN. Mergers that are
LIRGs exhibit a slightly higher incidence of AGN with respect to
the general IRAS-detected merger population and the control
sample. However, in mergers that are ULIRGs the ionization appears
to be dominated by the AGN. Six out of the ten ULIRGs in our
merger sample have a Seyfert-type AGN, while the other two are
classified as LINERs. This result appears consistent with the
recent literature which suggests that AGN become active in the
\emph{late} stages of a merger \citep{Sch07, Dar09b, Wild10, Kav11, Car12} and also with the findings of past studies (\citealp{San96,
Cle96, Ris00,Hop06, Cha07, You09, Tre09, Tre10, Iwa11}) which
suggest that AGN play an important role in the formation and
evolution of ULIRGs - at least 50$\%$ (and up to 75$\%$) of the
ULIRGs explored in past studies show an AGN signature. {Studies in Mid-IR in the past few years have found similar results, highlighted the more likely dominance SF in coalescing mergers \citep{Alo10, Per10}, showed a similar increase in AGN to SF activity in late stage mergers \citep{Pet11}, and also found an increase in composite SF/AGN objects \citep{Sti13}.}

\begin{figure*}
\centering
\begin{tabular}{c c c c}
\includegraphics[width=40mm]{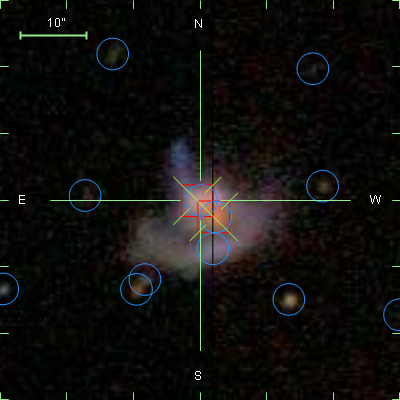} & \includegraphics[width=40mm]{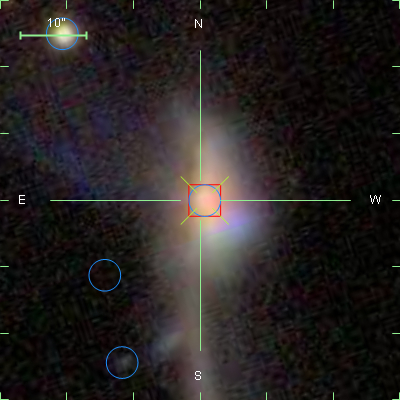} & \includegraphics[width=40mm]{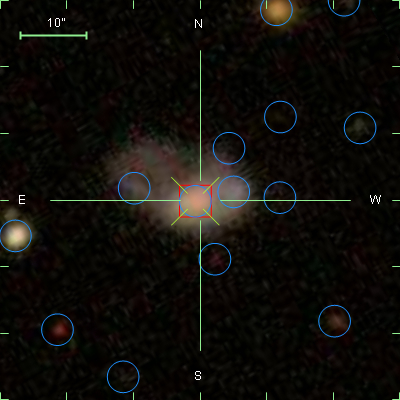}  & \includegraphics[width=40mm]{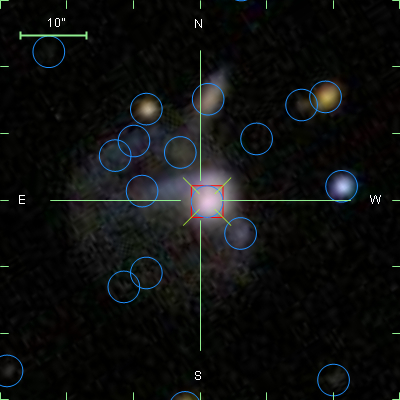}\\
\end{tabular}
\caption{\small SDSS images of 4 typical ULIRGs in our
sample: in order IRAS F12112+0305, Mrk 273, IRAS F15250+3608, IRAS F16133+2107.}\label{fig:ULIRG3}
\end{figure*}

\begin{figure*}
\centering
\begin{tabular}{c c}
\includegraphics[width=90mm]{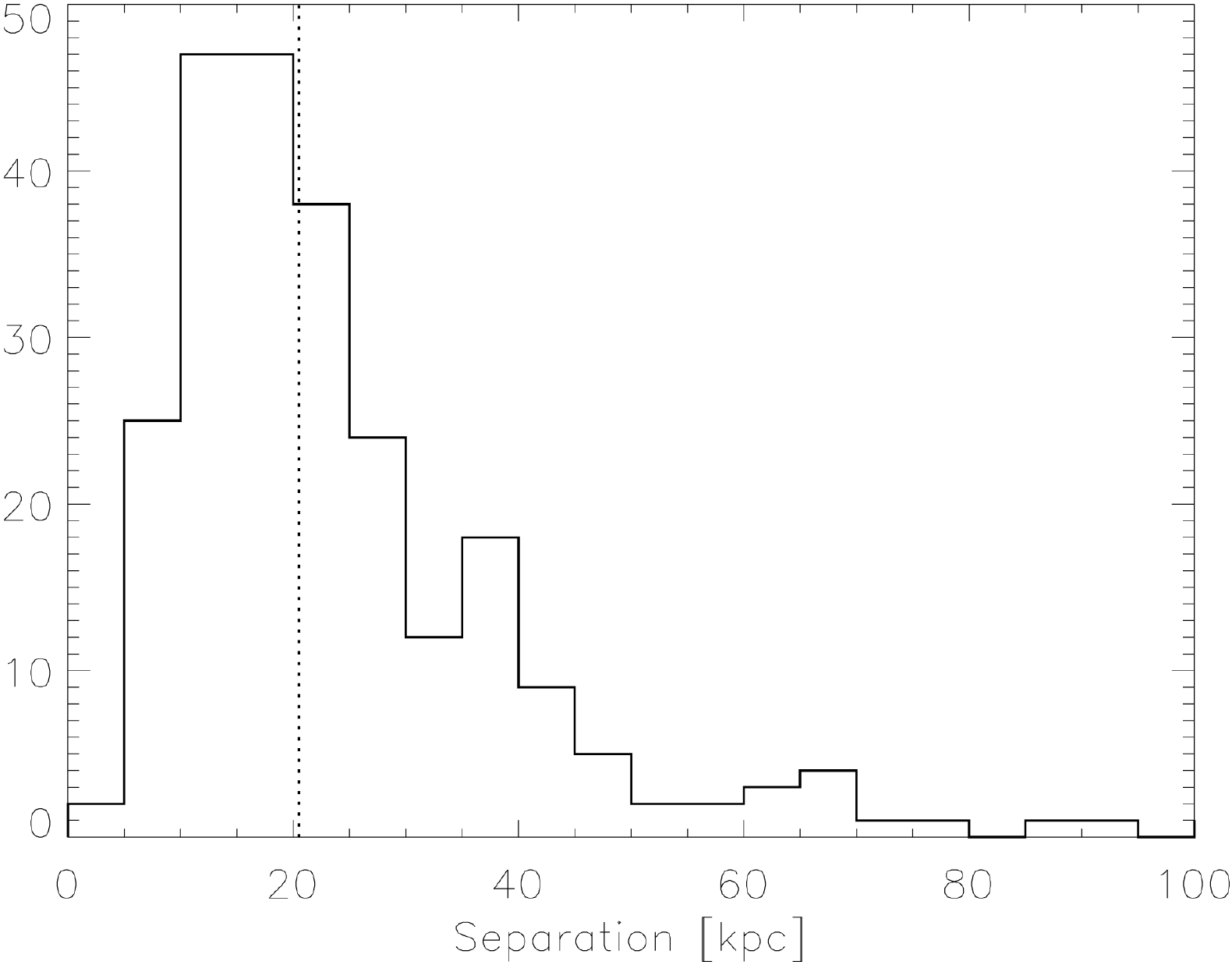} & \includegraphics[width=90mm]{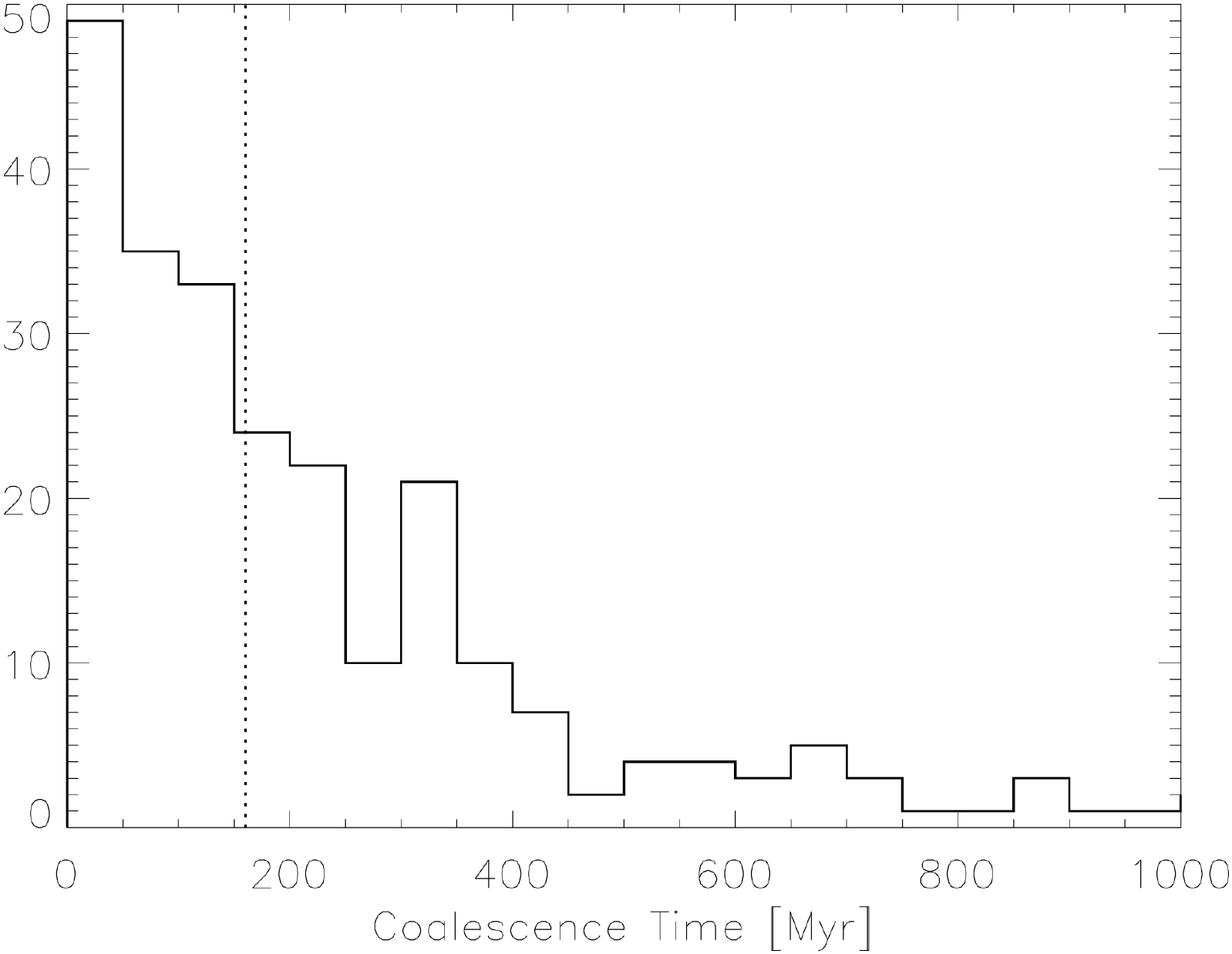}\\
\end{tabular}
\caption{\small Projected separation (left) and derived
coalescence timescales (right) of merging LIRGs in our sample. See
text for more details.}\label{fig:sep3}
\end{figure*}


{\section{Timescale of LIRG to ULIRG transformation along a merger
sequence} While our sample of ULIRGs is small, we find a clear
morphological segregation between LIRGs and ULIRGs in our sample.
In Figure \ref{fig:ULIRG3} we present typical SDSS images of
ULIRGs in our sample. Half the ULIRGs in our sample are in an
advanced merging state (the projected distance between their cores
is less than 4 kpc in all cases), while the other half have
already coalesced (i.e. it is not possible to resolve two cores in
the SDSS images). {The angular resolution of the SDSS is $\sim 1.3"$ \citep{Bra12}.
In comparison, $98\%$ of LIRGs have a core
separation greater than 4 kpc in the redshift limit of our sample (z=0.05 and z=0.1); 4 kpc equates to 4.1" and 2.2", respectively, much greater than the SDSS limit (left panel of Figure
\ref{fig:sep3}).} Assuming that an evolutionary transition occurs
from LIRG to ULIRG as the merger progresses and star formation
increases \citep{San96, Cle96, Das06}, it is instructive to
explore how quickly the LIRG-to-ULIRG transition is likely to take
place along the merger sequence. 

An estimate -- albeit crude --
can be derived for this coalescence timescale using the typical
separations of the LIRGs in our merger sample and the velocity
dispersion of the groups that they inhabit (which is a measure of
the typical relative velocities of objects in a group).

Keeping in mind that there may be a selection bias in the merger sample
against ULIRGs, we first check whether the ULIRGs in our merger
sample are representative of ULIRGs in general. We do so by
extracting all 121 ULIRGs in the IIFSCz at $z<0.1$ and inspecting
their SDSS images. We find that in this general ULIRG sample, more
than 90\% are indeed postmerger systems with a single core and
significant tidal debris, very similar to the ULIRGs in the merger
sample. The postmerger nature of ULIRGs in the Darg et al. sample
thus appears to be a typical characteristic of these systems, as
has been suggested by the past literature \citep{San96,Cle96,
Das06}.

In the bottom plot of Figure \ref{fig:sep3}, we estimate a
timescale for LIRGs to transform into ULIRGs along the merger
sequence. For most LIRGs we can only construct a projected
distance because fibre collisions mean that typically only one
progenitor has an SDSS redshift (so the line-of-sight separation
cannot be computed). Nevertheless, 32 merging LIRGs do have
redshifts for both progenitors. In these systems we compare the
projected separation and the true separation, to gauge the error
in the galaxy separations produced by using only projected
separations. We find that the true separations in this subsample
of LIRGs are only around $8\%$ larger than the projected
separations. In other words, projected separations seem to be
reasonably representative of true separations, at least in the
subsample of our merging LIRGs in which both progenitors have an
SDSS redshift. This is not unexpected, since these are merging
galaxies which, by definition, must be close to each other (i.e.
they must be virtually at the same redshift). Assuming this is
true for merging LIRGs in general, we assume the projected
separations to be reasonable estimates for the true separations in
all merging LIRGs in our sample.

To estimate a coalescence timescale we require an estimate for the
relative velocities of the individual galaxies. While we do not
have the information to measure either the relative transverse or
line-of-sight velocities (except in the 32 systems above where we
can only measure the relative line-of-sight velocities),we use
the mean velocity dispersion of SDSS groups ($258\pm7$) from \cite{Nur13}. 
{To estimate the timescale we use the following equation
\begin{equation}
T_{coal}=M_{ratio}\times 2\pi\times {r\over \sigma},
\end{equation}
where $r$ is the separation, $\sigma$ is the velocity dispersion, and $M_{ratio}$ the mass ratio of the two galaxies. The timescales are plotted in the right panel of Figure \ref{fig:sep3}}. From the
distribution of coalescence timescales, we estimate that the
timescale for the LIRG to ULIRG transformation is likely to be a
few hundred Myrs (160 $\pm 56$ Myr).{ Our empirically-determined
values and the range of the timescale appear to be in reasonably good agreement with previous surveys \citep{Mur96, Haa11}} as well as N-body/SPH merger simulations \citep[see e.g.][]{Tor12, Tor11}
which find coalescence timescales of $\sim 200$Myr. The low values
of the timescales indicate that \emph{merging LIRGs turn into
ULIRGs over reasonably short timescales, comparable to the
dynamical timescales of star formation in these systems.}


\section{Summary}
We have combined a sample of $\sim$3000 visually selected local
mergers from the Galaxy Zoo (GZ) project with the Imperial IRAS
Faint Source Redshift Catalogue (IIFSCz), to perform a large blind
far-infrared (FIR) study of the local merger population. 18\% of
the GZ mergers are detected by IRAS. Our results therefore
describe the brightest 18\% of merging systems in terms of FIR
luminosity (and therefore SFR).

The IRAS-detected mergers are overwhelmingly ($98\%$) comprised of
spiral-spiral systems, with a median FIR luminosity of $10^{11}
L_\odot$ and a median star formation rate of around $15 M_\odot
\;yr^{-1}$. The SFR of the IRAS-detected mergers are offset from
the general IRAS-detected galaxy population by 0.5 dex. {While the
merger SFR correlates with the total mass of the merging system,
we find very little dependence on the merger mass ratio. Indeed,
$65\%$ of the systems with LIRG luminosities (and possibly some
low-luminosity ULIRGs) are hosted by minor mergers (mass ratios
lower than $1:3$). This suggests that strong star formation (between 10 and $173 M_\odot yr^{-1}$) can be
triggered by the minor-merger process, in good agreement with the
recent literature.}

While they reside in low-density environments, the merger SFR does
not show a strong dependence on local density. An optical
emission-line ratio analysis indicates that the contribution of
AGN to the gas ionization increases with increasing FIR
luminosity. The AGN fraction increases dramatically for ULIRGs -
almost all ULIRGs show emission-line diagnostics indicative of an
AGN component, despite being strongly star-forming systems.

A particular characteristic of ULIRGs is that they are all in advance merger stage: either the nuclei are nearly coalesced (with a separation $<4$kpc) or they are already in a coalesced post merger phase. Assuming that the (non-coalesced) merging LIRGs
are the parent population for ULIRGs, we have estimated that
galaxies transition from being LIRGs to ULIRGs along the merger
sequence over a reasonably short timescale ($\sim160\pm56$ Myr).


\section*{Acknowledgments}
The authors thank the anonymous referee for comments and suggestions that greatly
improved this paper.
The data in this paper are the result of the efforts of the Galaxy Zoo volunteers, without whom none of this work would be possible. Their efforts are individually acknowledged at http://authors.galaxyzoo.org. Galaxy Zoo was supported by The Leverhulme Trust.
SK acknowledges a Senior Research Fellowship from Worcester
College Oxford.


\begin{small}
\bibliographystyle{bibtex/aa.bst}
\bibliography{bibliography}

\begin{thebibliography}{86}
\expandafter\ifx\csname natexlab\endcsname\relax\def\natexlab#1{#1}\fi

\bibitem[{{Adelman-McCarthy} {et~al.}(2008){Adelman-McCarthy}, {Ag{\"u}eros},
  {Allam}, {Allende Prieto}, {Anderson}, {Anderson}, {Annis}, {Bahcall},
  {Bailer-Jones}, {Baldry}, {Barentine}, {Bassett}, {Becker}, {Beers}, {Bell},
  {Berlind}, {Bernardi}, {Blanton}, {Bochanski}, {Boroski}, {Brinchmann},
  {Brinkmann}, {Brunner}, {Budav{\'a}ri}, {Carliles}, {Carr}, {Castander},
  {Cinabro}, {Cool}, {Covey}, {Csabai}, {Cunha}, {Davenport}, {Dilday}, {Doi},
  {Eisenstein}, {Evans}, {Fan}, {Finkbeiner}, {Friedman}, {Frieman},
  {Fukugita}, {G{\"a}nsicke}, {Gates}, {Gillespie}, {Glazebrook}, {Gray},
  {Grebel}, {Gunn}, {Gurbani}, {Hall}, {Harding}, {Harvanek}, {Hawley},
  {Hayes}, {Heckman}, {Hendry}, {Hindsley}, {Hirata}, {Hogan}, {Hogg}, {Hyde},
  {Ichikawa}, {Ivezi{\'c}}, {Jester}, {Johnson}, {Jorgensen}, {Juri{\'c}},
  {Kent}, {Kessler}, {Kleinman}, {Knapp}, {Kron}, {Krzesinski}, {Kuropatkin},
  {Lamb}, {Lampeitl}, {Lebedeva}, {Lee}, {Leger}, {L{\'e}pine}, {Lima}, {Lin},
  {Long}, {Loomis}, {Loveday}, {Lupton}, {Malanushenko}, {Malanushenko},
  {Mandelbaum}, {Margon}, {Marriner}, {Mart{\'{\i}}nez-Delgado}, {Matsubara},
  {McGehee}, {McKay}, {Meiksin}, {Morrison}, {Munn}, {Nakajima}, {Neilsen},
  {Newberg}, {Nichol}, {Nicinski}, {Nieto-Santisteban}, {Nitta}, {Okamura},
  {Owen}, {Oyaizu}, {Padmanabhan}, {Pan}, {Park}, {Peoples}, {Pier}, {Pope},
  {Purger}, {Raddick}, {Re Fiorentin}, {Richards}, {Richmond}, {Riess}, {Rix},
  {Rockosi}, {Sako}, {Schlegel}, {Schneider}, {Schreiber}, {Schwope}, {Seljak},
  {Sesar}, {Sheldon}, {Shimasaku}, {Sivarani}, {Smith}, {Snedden}, {Steinmetz},
  {Strauss}, {SubbaRao}, {Suto}, {Szalay}, {Szapudi}, {Szkody}, {Tegmark},
  {Thakar}, {Tremonti}, {Tucker}, {Uomoto}, {Vanden Berk}, {Vandenberg},
  {Vidrih}, {Vogeley}, {Voges}, {Vogt}, {Wadadekar}, {Weinberg}, {West},
  {White}, {Wilhite}, {Yanny}, {Yocum}, {York}, {Zehavi}, \& {Zucker}}]{Ade08}
{Adelman-McCarthy}, J.~K., {Ag{\"u}eros}, M.~A., {Allam}, S.~S., {et~al.} 2008,
  \apjs, 175, 297

\bibitem[{{Alonso-Herrero} {et~al.}(2010){Alonso-Herrero}, {Pereira-Santaella},
  {Rieke}, {Colina}, {Engelbracht}, {P{\'e}rez-Gonz{\'a}lez},
  {D{\'{\i}}az-Santos}, \& {Smith}}]{Alo10}
{Alonso-Herrero}, A., {Pereira-Santaella}, M., {Rieke}, G.~H., {et~al.} 2010,
  Advances in Space Research, 45, 99

\bibitem[{{Armus} {et~al.}(2009){Armus}, {Mazzarella}, {Evans}, {Surace},
  {Sanders}, {Iwasawa}, {Frayer}, {Howell}, {Chan}, {Petric}, {Vavilkin},
  {Kim}, {Haan}, {Inami}, {Murphy}, {Appleton}, {Barnes}, {Bothun}, {Bridge},
  {Charmandaris}, {Jensen}, {Kewley}, {Lord}, {Madore}, {Marshall},
  {Melbourne}, {Rich}, {Satyapal}, {Schulz}, {Spoon}, {Sturm}, {U}, {Veilleux},
  \& {Xu}}]{Arm09}
{Armus}, L., {Mazzarella}, J.~M., {Evans}, A.~S., {et~al.} 2009, \pasp, 121,
  559

\bibitem[{{Baldwin} {et~al.}(1981){Baldwin}, {Phillips}, \&
  {Terlevich}}]{BPT81}
{Baldwin}, J.~A., {Phillips}, M.~M., \& {Terlevich}, R. 1981, \pasp, 93, 5

\bibitem[{{Barnes} \& {Hernquist}(1991)}]{Bar91}
{Barnes}, J.~E. \& {Hernquist}, L.~E. 1991, \apjl, 370, L65

\bibitem[{{Blumenthal} {et~al.}(1984){Blumenthal}, {Faber}, {Primack}, \&
  {Rees}}]{Blu84}
{Blumenthal}, G.~R., {Faber}, S.~M., {Primack}, J.~R., \& {Rees}, M.~J. 1984,
  \nat, 311, 517

\bibitem[{{Bramich} \& {Freudling}(2012)}]{Bra12}
{Bramich}, D.~M. \& {Freudling}, W. 2012, \mnras, 424, 1584

\bibitem[{{Calzetti} {et~al.}(2000){Calzetti}, {Armus}, {Bohlin}, {Kinney},
  {Koornneef}, \& {Storchi-Bergmann}}]{Cal00}
{Calzetti}, D., {Armus}, L., {Bohlin}, R.~C., {et~al.} 2000, \apj, 533, 682

\bibitem[{{Carpineti} {et~al.}(2012){Carpineti}, {Kaviraj}, {Darg}, {Lintott},
  {Schawinski}, \& {Shabala}}]{Car12}
{Carpineti}, A., {Kaviraj}, S., {Darg}, D., {et~al.} 2012, \mnras, 420, 2139

\bibitem[{{Chakrabarti} {et~al.}(2007){Chakrabarti}, {Cox}, {Hernquist},
  {Hopkins}, {Robertson}, \& {Di Matteo}}]{Cha07}
{Chakrabarti}, S., {Cox}, T.~J., {Hernquist}, L., {et~al.} 2007, \apj, 658, 840

\bibitem[{{Chary} \& {Elbaz}(2001)}]{Cha01}
{Chary}, R. \& {Elbaz}, D. 2001, \apj, 556, 562

\bibitem[{{Clements} {et~al.}(1996){Clements}, {Sutherland}, {McMahon}, \&
  {Saunders}}]{Cle96}
{Clements}, D.~L., {Sutherland}, W.~J., {McMahon}, R.~G., \& {Saunders}, W.
  1996, \mnras, 279, 477

\bibitem[{{Cox} {et~al.}(2006{\natexlab{a}}){Cox}, {Dutta}, {Di Matteo},
  {Hernquist}, {Hopkins}, {Robertson}, \& {Springel}}]{Spr05}
{Cox}, T.~J., {Dutta}, S.~N., {Di Matteo}, T., {et~al.} 2006{\natexlab{a}},
  \apj, 650, 791

\bibitem[{{Cox} {et~al.}(2006{\natexlab{b}}){Cox}, {Dutta}, {Di Matteo},
  {Hernquist}, {Hopkins}, {Robertson}, \& {Springel}}]{Cox06}
{Cox}, T.~J., {Dutta}, S.~N., {Di Matteo}, T., {et~al.} 2006{\natexlab{b}},
  \apj, 650, 791

\bibitem[{{Darg} {et~al.}(2010{\natexlab{a}}){Darg}, {Kaviraj}, {Lintott},
  {Schawinski}, {Sarzi}, {Bamford}, {Silk}, {Andreescu}, {Murray}, {Nichol},
  {Raddick}, {Slosar}, {Szalay}, {Thomas}, \& {Vandenberg}}]{Dar09b}
{Darg}, D.~W., {Kaviraj}, S., {Lintott}, C.~J., {et~al.} 2010{\natexlab{a}},
  \mnras, 401, 1552

\bibitem[{{Darg} {et~al.}(2010{\natexlab{b}}){Darg}, {Kaviraj}, {Lintott},
  {Schawinski}, {Sarzi}, {Bamford}, {Silk}, {Proctor}, {Andreescu}, {Murray},
  {Nichol}, {Raddick}, {Slosar}, {Szalay}, {Thomas}, \& {Vandenberg}}]{Dar09a}
{Darg}, D.~W., {Kaviraj}, S., {Lintott}, C.~J., {et~al.} 2010{\natexlab{b}},
  \mnras, 401, 1043

\bibitem[{{Dasyra} {et~al.}(2006){Dasyra}, {Tacconi}, {Davies}, {Genzel},
  {Lutz}, {Naab}, {Burkert}, {Veilleux}, \& {Sanders}}]{Das06}
{Dasyra}, K.~M., {Tacconi}, L.~J., {Davies}, R.~I., {et~al.} 2006, \apj, 638,
  745

\bibitem[{{Di Matteo} {et~al.}(2005){Di Matteo}, {Springel}, \&
  {Hernquist}}]{Car05}
{Di Matteo}, T., {Springel}, V., \& {Hernquist}, L. 2005, \nat, 433, 604

\bibitem[{{Dole} {et~al.}(2006){Dole}, {Lagache}, {Puget}, {Caputi},
  {Fern{\'a}ndez-Conde}, {Le Floc'h}, {Papovich}, {P{\'e}rez-Gonz{\'a}lez},
  {Rieke}, \& {Blaylock}}]{Dol06}
{Dole}, H., {Lagache}, G., {Puget}, J.-L., {et~al.} 2006, \aap, 451, 417

\bibitem[{{Draine}(2003)}]{Dra03}
{Draine}, B.~T. 2003, \araa, 41, 241

\bibitem[{{Draine}(2009)}]{Dra09}
{Draine}, B.~T. 2009, in Astronomical Society of the Pacific Conference Series,
  Vol. 414, Cosmic Dust - Near and Far, ed. T.~{Henning}, E.~{Gr{\"u}n}, \&
  J.~{Steinacker}, 453

\bibitem[{{Efstathiou} {et~al.}(2002){Efstathiou}, {Moody}, {Peacock},
  {Percival}, {Baugh}, {Bland-Hawthorn}, {Bridges}, {Cannon}, {Cole},
  {Colless}, {Collins}, {Couch}, {Dalton}, {de Propris}, {Driver}, {Ellis},
  {Frenk}, {Glazebrook}, {Jackson}, {Lahav}, {Lewis}, {Lumsden}, {Maddox},
  {Norberg}, {Peterson}, {Sutherland}, \& {Taylor}}]{Efs02}
{Efstathiou}, G., {Moody}, S., {Peacock}, J.~A., {et~al.} 2002, \mnras, 330,
  L29

\bibitem[{{Elbaz} {et~al.}(2007){Elbaz}, {Daddi}, {Le Borgne}, {Dickinson},
  {Alexander}, {Chary}, {Starck}, {Brandt}, {Kitzbichler}, {MacDonald},
  {Nonino}, {Popesso}, {Stern}, \& {Vanzella}}]{Elb07}
{Elbaz}, D., {Daddi}, E., {Le Borgne}, D., {et~al.} 2007, \aap, 468, 33

\bibitem[{{Fixsen} {et~al.}(1998){Fixsen}, {Dwek}, {Mather}, {Bennett}, \&
  {Shafer}}]{Fix98}
{Fixsen}, D.~J., {Dwek}, E., {Mather}, J.~C., {Bennett}, C.~L., \& {Shafer},
  R.~A. 1998, \apj, 508, 123

\bibitem[{{Freedman} {et~al.}(2001){Freedman}, {Madore}, {Gibson}, {Ferrarese},
  {Kelson}, {Sakai}, {Mould}, {Kennicutt}, {Ford}, {Graham}, {Huchra},
  {Hughes}, {Illingworth}, {Macri}, \& {Stetson}}]{Fre01}
{Freedman}, W.~L., {Madore}, B.~F., {Gibson}, B.~K., {et~al.} 2001, \apj, 553,
  47

\bibitem[{{Haan} {et~al.}(2011){Haan}, {Surace}, {Armus}, {Evans}, {Howell},
  {Mazzarella}, {Kim}, {Vavilkin}, {Inami}, {Sanders}, {Petric}, {Bridge},
  {Melbourne}, {Charmandaris}, {Diaz-Santos}, {Murphy}, {U}, {Stierwalt}, \&
  {Marshall}}]{Haa11}
{Haan}, S., {Surace}, J.~A., {Armus}, L., {et~al.} 2011, \aj, 141, 100

\bibitem[{{Hauser} \& {Dwek}(2001)}]{Hau01}
{Hauser}, M.~G. \& {Dwek}, E. 2001, \araa, 39, 249

\bibitem[{{Hopkins} {et~al.}(2006){Hopkins}, {Somerville}, {Hernquist}, {Cox},
  {Robertson}, \& {Li}}]{Hop06}
{Hopkins}, P.~F., {Somerville}, R.~S., {Hernquist}, L., {et~al.} 2006, \apj,
  652, 864

\bibitem[{{Hutchings} \& {Neff}(1987)}]{Hut87}
{Hutchings}, J.~B. \& {Neff}, S.~G. 1987, \aj, 93, 14

\bibitem[{{Hwang} {et~al.}(2010){Hwang}, {Elbaz}, {Lee}, {Jeong}, {Park},
  {Lee}, \& {Lee}}]{Hwa10}
{Hwang}, H.~S., {Elbaz}, D., {Lee}, J.~C., {et~al.} 2010, \aap, 522, A33

\bibitem[{{Iwasawa} {et~al.}(2011){Iwasawa}, {Sanders}, {Teng}, {U}, {Armus},
  {Evans}, {Howell}, {Komossa}, {Mazzarella}, {Petric}, {Surace}, {Vavilkin},
  {Veilleux}, \& {Trentham}}]{Iwa11}
{Iwasawa}, K., {Sanders}, D.~B., {Teng}, S.~H., {et~al.} 2011, \aap, 529, A106

\bibitem[{{Kauffmann} {et~al.}(2003){Kauffmann}, {Heckman}, {Tremonti},
  {Brinchmann}, {Charlot}, {White}, {Ridgway}, {Brinkmann}, {Fukugita}, {Hall},
  {Ivezi{\'c}}, {Richards}, \& {Schneider}}]{Kau03}
{Kauffmann}, G., {Heckman}, T.~M., {Tremonti}, C., {et~al.} 2003, \mnras, 346,
  1055

\bibitem[{{Kaviraj}(2014{\natexlab{a}})}]{Kav14b}
{Kaviraj}, S. 2014{\natexlab{a}}, \mnras, 440, 2944

\bibitem[{{Kaviraj}(2014{\natexlab{b}})}]{Kav14}
{Kaviraj}, S. 2014{\natexlab{b}}, \mnras, 437, L41

\bibitem[{{Kaviraj} {et~al.}(2009){Kaviraj}, {Peirani}, {Khochfar}, {Silk}, \&
  {Kay}}]{Kav09}
{Kaviraj}, S., {Peirani}, S., {Khochfar}, S., {Silk}, J., \& {Kay}, S. 2009,
  \mnras, 394, 1713

\bibitem[{{Kaviraj} {et~al.}(2011){Kaviraj}, {Schawinski}, {Silk}, \&
  {Shabala}}]{Kav11}
{Kaviraj}, S., {Schawinski}, K., {Silk}, J., \& {Shabala}, S.~S. 2011, \mnras,
  415, 3798

\bibitem[{{Kennicutt}(1998)}]{Ken98}
{Kennicutt}, Jr., R.~C. 1998, \araa, 36, 189

\bibitem[{{Kewley} {et~al.}(2006){Kewley}, {Groves}, {Kauffmann}, \&
  {Heckman}}]{Kew06}
{Kewley}, L.~J., {Groves}, B., {Kauffmann}, G., \& {Heckman}, T. 2006, \mnras,
  372, 961

\bibitem[{{Kleinmann} \& {Keel}(1987)}]{Kle87}
{Kleinmann}, S. \& {Keel}, W. 1987, Star Formation in Galaxies (C.J.
  Lonsdale-Persson, Washington DC: US Govt. Print. Off.)

\bibitem[{{Lintott} {et~al.}(2008){Lintott}, {Schawinski}, {Slosar}, {Land},
  {Bamford}, {Thomas}, {Raddick}, {Nichol}, {Szalay}, {Andreescu}, {Murray}, \&
  {Vandenberg}}]{Lin08}
{Lintott}, C.~J., {Schawinski}, K., {Slosar}, A., {et~al.} 2008, \mnras, 389,
  1179

\bibitem[{{Maraston}(1998)}]{Mara98}
{Maraston}, C. 1998, \mnras, 300, 872

\bibitem[{{Maraston}(2005)}]{Mara05}
{Maraston}, C. 2005, \mnras, 362, 799

\bibitem[{{Mihos} \& {Hernquist}(1996{\natexlab{a}})}]{Mih96}
{Mihos}, J.~C. \& {Hernquist}, L. 1996{\natexlab{a}}, \apj, 464, 641

\bibitem[{{Mihos} \& {Hernquist}(1996{\natexlab{b}})}]{Bar96}
{Mihos}, J.~C. \& {Hernquist}, L. 1996{\natexlab{b}}, \apj, 464, 641

\bibitem[{{Moshir} {et~al.}(1992){Moshir}, {Kopman}, \& {Conrow}}]{Mos92}
{Moshir}, M., {Kopman}, G., \& {Conrow}, T.~A.~O. 1992, {IRAS Faint Source
  Survey, Explanatory supplement version 2} (California Institute of
  Technology)

\bibitem[{{Murphy} {et~al.}(1996){Murphy}, {Armus}, {Matthews}, {Soifer},
  {Mazzarella}, {Shupe}, {Strauss}, \& {Neugebauer}}]{Mur96}
{Murphy}, Jr., T.~W., {Armus}, L., {Matthews}, K., {et~al.} 1996, \aj, 111,
  1025

\bibitem[{{Neugebauer} {et~al.}(1984{\natexlab{a}}){Neugebauer}, {Habing}, {van
  Duinen}, {Aumann}, {Baud}, {Beichman}, {Beintema}, {Boggess}, {Clegg}, {de
  Jong}, {Emerson}, {Gautier}, {Gillett}, {Harris}, {Hauser}, {Houck},
  {Jennings}, {Low}, {Marsden}, {Miley}, {Olnon}, {Pottasch}, {Raimond},
  {Rowan-Robinson}, {Soifer}, {Walker}, {Wesselius}, \& {Young}}]{IRAS}
{Neugebauer}, G., {Habing}, H.~J., {van Duinen}, R., {et~al.}
  1984{\natexlab{a}}, \apjl, 278, L1

\bibitem[{{Neugebauer} {et~al.}(1984{\natexlab{b}}){Neugebauer}, {Soifer},
  {Miley}, {Habing}, {Young}, {Low}, {Beichman}, {Clegg}, {Harris}, \&
  {Rowan-Robinson}}]{Neu84}
{Neugebauer}, G., {Soifer}, B.~T., {Miley}, G., {et~al.} 1984{\natexlab{b}},
  \apjl, 278, L83

\bibitem[{{Nurmi} {et~al.}(2013){Nurmi}, {Hein{\"a}m{\"a}ki}, {Sepp}, {Tago},
  {Saar}, {Gramann}, {Einasto}, {Tempel}, \& {Einasto}}]{Nur13}
{Nurmi}, P., {Hein{\"a}m{\"a}ki}, P., {Sepp}, T., {et~al.} 2013, \mnras, 436,
  380

\bibitem[{{Pei} {et~al.}(1999){Pei}, {Fall}, \& {Hauser}}]{Pei99}
{Pei}, Y.~C., {Fall}, S.~M., \& {Hauser}, M.~G. 1999, \apj, 522, 604

\bibitem[{{Pereira-Santaella} {et~al.}(2010){Pereira-Santaella},
  {Alonso-Herrero}, {Rieke}, {Colina}, {D{\'{\i}}az-Santos}, {Smith},
  {P{\'e}rez-Gonz{\'a}lez}, \& {Engelbracht}}]{Per10}
{Pereira-Santaella}, M., {Alonso-Herrero}, A., {Rieke}, G.~H., {et~al.} 2010,
  \apjs, 188, 447

\bibitem[{{Petric} {et~al.}(2011){Petric}, {Armus}, {Howell}, {Chan},
  {Mazzarella}, {Evans}, {Surace}, {Sanders}, {Appleton}, {Charmandaris},
  {D{\'{\i}}az-Santos}, {Frayer}, {Haan}, {Inami}, {Iwasawa}, {Kim}, {Madore},
  {Marshall}, {Spoon}, {Stierwalt}, {Sturm}, {U}, {Vavilkin}, \&
  {Veilleux}}]{Pet11}
{Petric}, A.~O., {Armus}, L., {Howell}, J., {et~al.} 2011, \apj, 730, 28

\bibitem[{{Pryke} {et~al.}(2002){Pryke}, {Halverson}, {Leitch}, {Kovac},
  {Carlstrom}, {Holzapfel}, \& {Dragovan}}]{Pry02}
{Pryke}, C., {Halverson}, N.~W., {Leitch}, E.~M., {et~al.} 2002, \apj, 568, 46

\bibitem[{{Puget} {et~al.}(1996){Puget}, {Abergel}, {Bernard}, {Boulanger},
  {Burton}, {Desert}, \& {Hartmann}}]{Pug96}
{Puget}, J.-L., {Abergel}, A., {Bernard}, J.-P., {et~al.} 1996, \aap, 308, L5

\bibitem[{{Risaliti} {et~al.}(2000){Risaliti}, {Gilli}, {Maiolino}, \&
  {Salvati}}]{Ris00}
{Risaliti}, G., {Gilli}, R., {Maiolino}, R., \& {Salvati}, M. 2000, \aap, 357,
  13

\bibitem[{{Rowan-Robinson} {et~al.}(2008){Rowan-Robinson}, {Babbedge},
  {Oliver}, {Trichas}, {Berta}, {Lonsdale}, {Smith}, {Shupe}, {Surace},
  {Arnouts}, {Ilbert}, {Le F{\'e}vre}, {Afonso-Luis}, {Perez-Fournon},
  {Hatziminaoglou}, {Polletta}, {Farrah}, \& {Vaccari}}]{RR08}
{Rowan-Robinson}, M., {Babbedge}, T., {Oliver}, S., {et~al.} 2008, \mnras, 386,
  697

\bibitem[{{Salpeter}(1955)}]{Sal55}
{Salpeter}, E.~E. 1955, \apj, 121, 161

\bibitem[{{Sanders} \& {Mirabel}(1996)}]{San96}
{Sanders}, D.~B. \& {Mirabel}, I.~F. 1996, \araa, 34, 749

\bibitem[{{Sanders} {et~al.}(1987){Sanders}, {Scoville}, {Soifer}, {Young}, \&
  {Danielson}}]{San87}
{Sanders}, D.~B., {Scoville}, N.~Z., {Soifer}, B.~T., {Young}, J.~S., \&
  {Danielson}, G.~E. 1987, \apjl, 312, L5

\bibitem[{{Sanders} {et~al.}(1988{\natexlab{a}}){Sanders}, {Soifer}, {Elias},
  {Madore}, {Matthews}, {Neugebauer}, \& {Scoville}}]{San88}
{Sanders}, D.~B., {Soifer}, B.~T., {Elias}, J.~H., {et~al.} 1988{\natexlab{a}},
  \apj, 325, 74

\bibitem[{{Sanders} {et~al.}(1988{\natexlab{b}}){Sanders}, {Soifer}, {Elias},
  {Neugebauer}, \& {Matthews}}]{San88b}
{Sanders}, D.~B., {Soifer}, B.~T., {Elias}, J.~H., {Neugebauer}, G., \&
  {Matthews}, K. 1988{\natexlab{b}}, \apjl, 328, L35

\bibitem[{{Sarzi} {et~al.}(2006){Sarzi}, {Falc{\'o}n-Barroso}, {Davies},
  {Bacon}, {Bureau}, {Cappellari}, {de Zeeuw}, {Emsellem}, {Fathi},
  {Krajnovi{\'c}}, {Kuntschner}, {McDermid}, \& {Peletier}}]{Sar06}
{Sarzi}, M., {Falc{\'o}n-Barroso}, J., {Davies}, R.~L., {et~al.} 2006, \mnras,
  366, 1151

\bibitem[{{Schawinski} {et~al.}(2007){Schawinski}, {Kaviraj}, {Khochfar},
  {Yoon}, {Yi}, {Deharveng}, {Boselli}, {Barlow}, {Conrow}, {Forster},
  {Friedman}, {Martin}, {Morrissey}, {Neff}, {Schiminovich}, {Seibert},
  {Small}, {Wyder}, {Bianchi}, {Donas}, {Heckman}, {Lee}, {Madore}, {Milliard},
  {Rich}, \& {Szalay}}]{Sch07}
{Schawinski}, K., {Kaviraj}, S., {Khochfar}, S., {et~al.} 2007, \apjs, 173, 512

\bibitem[{{Schawinski} {et~al.}(2009){Schawinski}, {Virani}, {Simmons}, {Urry},
  {Treister}, {Kaviraj}, \& {Kushkuley}}]{Sch09}
{Schawinski}, K., {Virani}, S., {Simmons}, B., {et~al.} 2009, \apjl, 692, L19

\bibitem[{{Scudder} {et~al.}(2012){Scudder}, {Ellison}, {Torrey}, {Patton}, \&
  {Mendel}}]{Scu12}
{Scudder}, J.~M., {Ellison}, S.~L., {Torrey}, P., {Patton}, D.~R., \& {Mendel},
  J.~T. 2012, \mnras, 426, 549

\bibitem[{{Searle} \& {Zinn}(1978)}]{Sea78}
{Searle}, L. \& {Zinn}, R. 1978, \apj, 225, 357

\bibitem[{{Shu} {et~al.}(1987){Shu}, {Adams}, \& {Lizano}}]{Shu87}
{Shu}, F.~H., {Adams}, F.~C., \& {Lizano}, S. 1987, \araa, 25, 23

\bibitem[{{Soifer} {et~al.}(1984){Soifer}, {Rowan-Robinson}, {Houck}, {de
  Jong}, {Neugebauer}, {Aumann}, {Beichman}, {Boggess}, {Clegg}, {Emerson},
  {Gillett}, {Habing}, {Hauser}, {Low}, {Miley}, \& {Young}}]{Soi84}
{Soifer}, B.~T., {Rowan-Robinson}, M., {Houck}, J.~R., {et~al.} 1984, \apjl,
  278, L71

\bibitem[{{Spergel} {et~al.}(2007){Spergel}, {Bean}, {Dor{\'e}}, {Nolta},
  {Bennett}, {Dunkley}, {Hinshaw}, {Jarosik}, {Komatsu}, {Page}, {Peiris},
  {Verde}, {Halpern}, {Hill}, {Kogut}, {Limon}, {Meyer}, {Odegard}, {Tucker},
  {Weiland}, {Wollack}, \& {Wright}}]{Spe07}
{Spergel}, D.~N., {Bean}, R., {Dor{\'e}}, O., {et~al.} 2007, \apjs, 170, 377

\bibitem[{{Spitzer}(1978)}]{Spi78}
{Spitzer}, L. 1978, {Physical processes in the interstellar medium} (Wiley-VCH)

\bibitem[{{Stierwalt} {et~al.}(2013){Stierwalt}, {Armus}, {Surace}, {Inami},
  {Petric}, {Diaz-Santos}, {Haan}, {Charmandaris}, {Howell}, {Kim}, {Marshall},
  {Mazzarella}, {Spoon}, {Veilleux}, {Evans}, {Sanders}, {Appleton}, {Bothun},
  {Bridge}, {Chan}, {Frayer}, {Iwasawa}, {Kewley}, {Lord}, {Madore},
  {Melbourne}, {Murphy}, {Rich}, {Schulz}, {Sturm}, {Vavilkin}, \&
  {Xu}}]{Sti13}
{Stierwalt}, S., {Armus}, L., {Surace}, J.~A., {et~al.} 2013, \apjs, 206, 1

\bibitem[{{Toomre}(1977)}]{Too77}
{Toomre}, A. 1977, in Evolution of Galaxies and Stellar Populations, ed. B.~M.
  {Tinsley} \& R.~B.~G. {Larson}, D.~Campbell, 401

\bibitem[{{Torrey} {et~al.}(2012{\natexlab{a}}){Torrey}, {Cox}, {Kewley}, \&
  {Hernquist}}]{Tor12}
{Torrey}, P., {Cox}, T.~J., {Kewley}, L., \& {Hernquist}, L.
  2012{\natexlab{a}}, \apj, 746, 108

\bibitem[{{Torrey} {et~al.}(2012{\natexlab{b}}){Torrey}, {Vogelsberger},
  {Sijacki}, {Springel}, \& {Hernquist}}]{Tor11}
{Torrey}, P., {Vogelsberger}, M., {Sijacki}, D., {Springel}, V., \&
  {Hernquist}, L. 2012{\natexlab{b}}, \mnras, 427, 2224

\bibitem[{{Treister} {et~al.}(2009){Treister}, {Cardamone}, {Schawinski},
  {Urry}, {Gawiser}, {Virani}, {Lira}, {Kartaltepe}, {Damen}, {Taylor}, {Le
  Floc'h}, {Justham}, \& {Koekemoer}}]{Tre09}
{Treister}, E., {Cardamone}, C.~N., {Schawinski}, K., {et~al.} 2009, \apj, 706,
  535

\bibitem[{{Treister} {et~al.}(2010){Treister}, {Natarajan}, {Sanders}, {Urry},
  {Schawinski}, \& {Kartaltepe}}]{Tre10}
{Treister}, E., {Natarajan}, P., {Sanders}, D.~B., {et~al.} 2010, Science, 328,
  600

\bibitem[{{Vader} \& {Simon}(1987)}]{Vad87}
{Vader}, J.~P. \& {Simon}, M. 1987, \nat, 327, 304

\bibitem[{{Veilleux} \& {Osterbrock}(1987)}]{Vei87}
{Veilleux}, S. \& {Osterbrock}, D.~E. 1987, in NASA Conference Publication,
  Vol. 2466, NASA Conference Publication, ed. C.~J. {Lonsdale Persson},
  737--740

\bibitem[{{Wang} \& {Rowan-Robinson}(2009)}]{Wan09}
{Wang}, L. \& {Rowan-Robinson}, M. 2009, \mnras, 398, 109

\bibitem[{{White} \& {Rees}(1978)}]{Whi78}
{White}, S.~D.~M. \& {Rees}, M.~J. 1978, \mnras, 183, 341

\bibitem[{{Wild} {et~al.}(2010){Wild}, {Heckman}, \& {Charlot}}]{Wild10}
{Wild}, V., {Heckman}, T., \& {Charlot}, S. 2010, \mnras, 405, 933

\bibitem[{{Willett} {et~al.}(2015){Willett}, {Schawinski}, {Simmons},
  {Masters}, {Skibba}, {Kaviraj}, {Melvin}, {Wong}, {Nichol}, {Cheung},
  {Lintott}, \& {Fortson}}]{Will15}
{Willett}, K.~W., {Schawinski}, K., {Simmons}, B.~D., {et~al.} 2015, ArXiv
  e-prints

\bibitem[{{Yang} {et~al.}(2007){Yang}, {Mo}, {van den Bosch}, {Pasquali}, {Li},
  \& {Barden}}]{Yang07}
{Yang}, X., {Mo}, H.~J., {van den Bosch}, F.~C., {et~al.} 2007, \apj, 671, 153

\bibitem[{{York} {et~al.}(2000){York}, {Adelman}, {Anderson}, {Anderson},
  {Annis}, {Bahcall}, {Bakken}, {Barkhouser}, {Bastian}, {Berman}, {Boroski},
  {Bracker}, {Briegel}, {Briggs}, {Brinkmann}, {Brunner}, {Burles}, {Carey},
  {Carr}, {Castander}, {Chen}, {Colestock}, {Connolly}, {Crocker}, {Csabai},
  {Czarapata}, {Davis}, {Doi}, {Dombeck}, {Eisenstein}, {Ellman}, {Elms},
  {Evans}, {Fan}, {Federwitz}, {Fiscelli}, {Friedman}, {Frieman}, {Fukugita},
  {Gillespie}, {Gunn}, {Gurbani}, {de Haas}, {Haldeman}, {Harris}, {Hayes},
  {Heckman}, {Hennessy}, {Hindsley}, {Holm}, {Holmgren}, {Huang}, {Hull},
  {Husby}, {Ichikawa}, {Ichikawa}, {Ivezi{\'c}}, {Kent}, {Kim}, {Kinney},
  {Klaene}, {Kleinman}, {Kleinman}, {Knapp}, {Korienek}, {Kron}, {Kunszt},
  {Lamb}, {Lee}, {Leger}, {Limmongkol}, {Lindenmeyer}, {Long}, {Loomis},
  {Loveday}, {Lucinio}, {Lupton}, {MacKinnon}, {Mannery}, {Mantsch}, {Margon},
  {McGehee}, {McKay}, {Meiksin}, {Merelli}, {Monet}, {Munn}, {Narayanan},
  {Nash}, {Neilsen}, {Neswold}, {Newberg}, {Nichol}, {Nicinski}, {Nonino},
  {Okada}, {Okamura}, {Ostriker}, {Owen}, {Pauls}, {Peoples}, {Peterson},
  {Petravick}, {Pier}, {Pope}, {Pordes}, {Prosapio}, {Rechenmacher}, {Quinn},
  {Richards}, {Richmond}, {Rivetta}, {Rockosi}, {Ruthmansdorfer}, {Sandford},
  {Schlegel}, {Schneider}, {Sekiguchi}, {Sergey}, {Shimasaku}, {Siegmund},
  {Smee}, {Smith}, {Snedden}, {Stone}, {Stoughton}, {Strauss}, {Stubbs},
  {SubbaRao}, {Szalay}, {Szapudi}, {Szokoly}, {Thakar}, {Tremonti}, {Tucker},
  {Uomoto}, {Vanden Berk}, {Vogeley}, {Waddell}, {Wang}, {Watanabe},
  {Weinberg}, {Yanny}, {Yasuda}, \& {SDSS Collaboration}}]{Yor00}
{York}, D.~G., {Adelman}, J., {Anderson}, Jr., J.~E., {et~al.} 2000, \aj, 120,
  1579

\bibitem[{{Younger} {et~al.}(2009){Younger}, {Hayward}, {Narayanan}, {Cox},
  {Hernquist}, \& {Jonsson}}]{You09}
{Younger}, J.~D., {Hayward}, C.~C., {Narayanan}, D., {et~al.} 2009, \mnras,
  396, L66

\bibitem[{{Zubko} {et~al.}(2004){Zubko}, {Dwek}, \& {Arendt}}]{Zub04}
{Zubko}, V., {Dwek}, E., \& {Arendt}, R.~G. 2004, \apjs, 152, 211

\end{thebibliography}
\end{small}

\end{document}